\documentclass{article}

\usepackage[preprint, nonatbib]{neurips_2023}
\usepackage[utf8]{inputenc} % allow utf-8 input
\usepackage[T1]{fontenc}    % use 8-bit T1 fonts
\usepackage{hyperref}       % hyperlinks
\usepackage{url}            % simple URL typesetting
\usepackage{booktabs}       % professional-quality tables
\usepackage{amsfonts}       % blackboard math symbols
\usepackage{nicefrac}       % compact symbols for 1/2, etc.
\usepackage{microtype}      % microtypography
\usepackage{xcolor}         % colors
\usepackage{graphicx}
\usepackage{subcaption}

\title{Data Efficiency for Large Recommendation Models}

\author{
  Kshitij Jain \\
  Google \\
  \texttt{kshitijj@google.com} \\ \And
Jingru Xie \\
Google\\
\texttt{jingrux@google.com} \\
\And
Kevin Regan \\
 Google \\
 \texttt{kevinregan@google.com} \\ 
\And
Cheng Chen \\  Google  \And
Jie Han \\  Google  \And
Steve Li \\  Google \And
Zhuoshu Li  \\  Google \And
Todd Phillips \\  Google  \And
Myles Sussman \\  Google \And
Matt Troup \\  Google  \And
Angel Yu \\  Google  \And
Jia Zhuo \\  Google 
}

\begin{document}

\maketitle

\begin{abstract}
Large recommendation models (LRMs) are fundamental to the multi-billion dollar online advertising industry, processing massive datasets of hundreds of billions of examples before transitioning to continuous online training to adapt to rapidly changing user behavior \cite{anil2022factory}. The massive scale of data directly impacts both computational costs and the speed at which new methods can be evaluated (R\&D velocity).

This paper presents actionable principles and high-level frameworks to guide practitioners in optimizing training data requirements. These strategies have been successfully deployed in Google's largest Ads CTR prediction models \cite{anil2022factory, mcmahan2013ad} and are broadly applicable beyond LRMs. We outline the concept of \emph{data convergence}, describe methods to accelerate this convergence, and finally, detail how to optimally balance training data volume with model size.
\end{abstract}

\section{Introduction}

Recent years have seen an exponential growth in the scale of model size, training data, and computational resources employed for the training of large recommendation models (LRMs). We have observed that the benefits from simply scaling these factors further have begun to plateau, highlighting the need for a shift towards optimizing efficiency and maximizing return on investment. One critical area ripe for optimization is the volume of training data, as LRMs often rely on datasets comprising hundreds of billions of examples \cite{anil2022factory}. 

The training of LRMs usually consists of 1) initial training where the model is trained to converge on hundreds of billions of samples in a single sequential pass over chronologically ordered historical data and 2) continuous online training, where the model continues learning with new data coming in at the rate of billions examples per day \cite{mcmahan2013ad}. Initial training is important to guarantee model convergence, while continuous online training is important for the model to stay fresh. Both of these training components are crucial for stable online  A/B testing in live traffic \cite{43887}. In this paper, we delve into strategies to enhance data efficiency in both initial training and continuous online training stage. We explored methods including downsampling and continuous distillation to minimize training data while maintaining model convergence and performance. We further investigated the tradeoff between data volume and model size in the context of improving overall resource efficiency. Through a comprehensive exploration of these factors, this paper aims to provide a roadmap for achieving greater efficiency in the development and deployment of large recommendation models.

\section{Data Convergence}

In our setting, increasing the historical training dataset size eventually yields diminishing accuracy returns. When launching model improvements, business metrics are verified during a fixed period of A/B testing with live traffic. It is important that model accuracy is stable during this period—i.e., the model has converged and accuracy does not significantly improve (or degrade) over time\footnote{Accuracy metrics like logloss are sensitive to fluctuations in click-through rates due to user behavior; we mitigate this by assessing convergence relative to a well-established baseline model that has seen substantially more data and for which we are much more certain of convergence}. We refer to such a point as \emph{convergence point}, where the model is converged w.r.t data. 

\begin{figure}
  \includegraphics[width=0.8\linewidth]{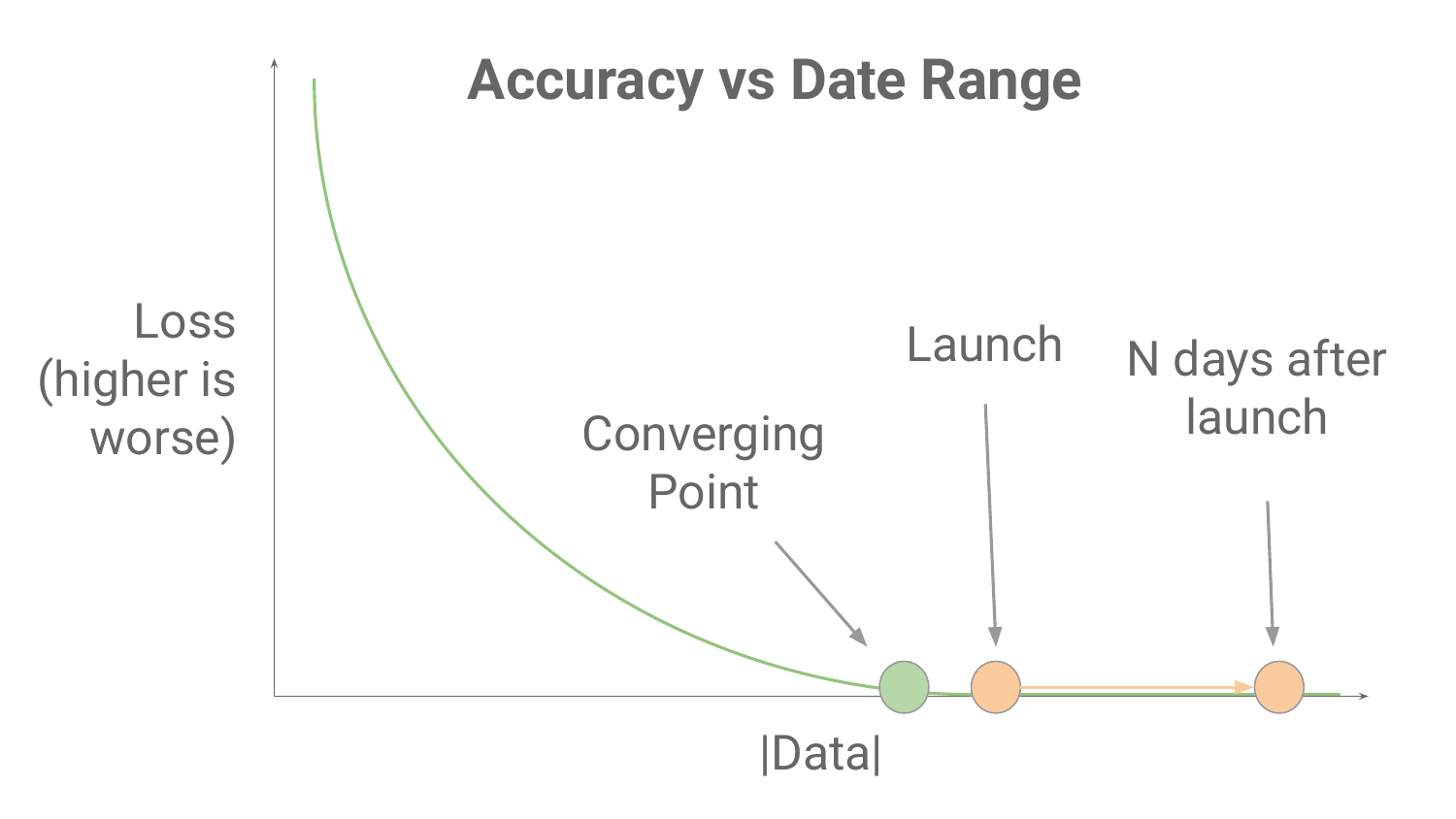}
  \caption{Model's accuracy as dataset grows highlighting model's \emph{convergence point} w.r.t data}
  \label{fig:SDR}
\end{figure}

To reduce training time, periodic dataset pruning post-convergence is effective (Fig. \ref{fig:SDR}). Specifically, we advance the start date of the historical dataset by the number of days since the last launch, maintaining convergence (assuming no major model changes). This, coupled with tuning dataset hyper-parameters (e.g., downsampling rate and duration, Sec. \ref{sec:downsampling}), substantially reduces training time.

\section{Downsampling}
\label{sec:downsampling}

LRMs for ad click prediction often face label imbalance, with positive (clicked) events being significantly outnumbered by negative (unclicked) events (e.g., a ratio of 1:250 \cite{10066174}).  Substantial downsampling of negative examples is feasible without compromising model accuracy \cite{anil2022factory}. Importance re-weighting, where kept examples are up-weighted by the inverse sampling rate, is used to mitigate prediction bias.

Beyond basic negative example removal, past work has shown success in downsampling examples with low logistic loss and low likelihood of user exposure \cite{anil2022factory}. Additionally, using uncertainty estimates derived from feature occurrence counts \cite{koenker1978regression} effectively focusing training on examples where the model is less confident, further reducing data requirements. 

\subsection{Continuous downsampling}

\begin{figure}[ht]
  \includegraphics[width=0.8\linewidth]{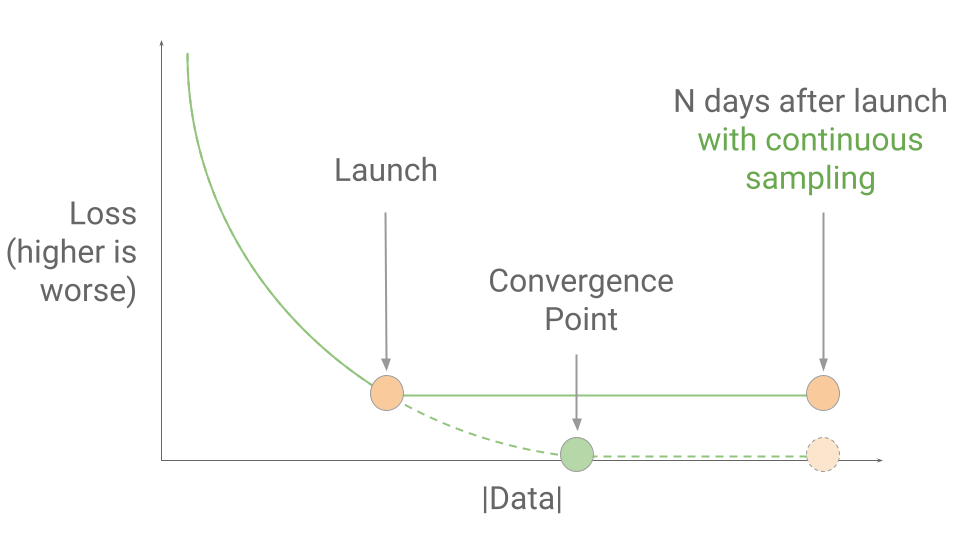}
  \caption{Moving the model's \emph{convergence point} using continuous sampling.}
  \label{fig:caughtup-SDR}
\end{figure}

A common downsampling strategy uses a \emph{cutoff}, after which downsampling stops. This enables aggressive initial downsampling, potentially impacting early accuracy, with recovery post-cutoff. However for certain models, accuracy often continues improving post-cutoff, increasing the data volume needed for convergence.

Instead, we propose \emph{continuous downsampling}, where downsampling persists throughout training, including online training. This necessitates "fresh" information for selecting examples to downsample alongside online data, introducing complexity. Downsampling rates can be tuned to balance accuracy and cost. With a consistent downsampling algorithm, any quality reduction is constant over time, effectively accelerating convergence, as shown in Fig. \ref{fig:chart_cud}. Such a reduction comes at the cost of a small constant accuracy hit.  

\section{Continuous Distillation}

% Note on estimates of cutover times:
% PQAL measurements were made in March 2024, G112 trained on ~365 days of data at that point. 
% 75 days before march 2024 is 75/365 ~= 80% of the way through pre-training
% 135 days before march 2024  is 135/365 ~= 60% of the way through pre-training
% 

Knowledge distillation \cite{hinton2015distilling} from a larger teacher model is a common technique to improve model accuracy. Traditionally, distillation was halted at a fixed point during training on historical data. We contrast this with continuous distillation, where distillation extends into online training, allowing both teacher and student models to learn concurrently from new data. 

\begin{figure}[ht]
  \includegraphics[width=0.9\linewidth]{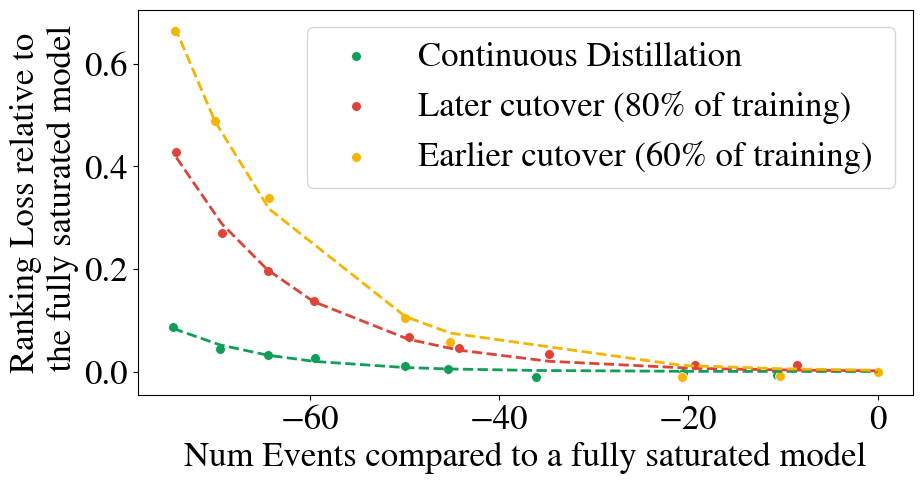}
  \caption{Continuous distillation accelerates model convergence. }
  \label{fig:chart_cud}
\end{figure}

To assess the efficacy of continuous distillation, we trained three model sets: (a) continuous distillation models, (b) later cutover models with a cutover date ~80\% through our production model's historical training period, and (c) early cutover models with a cutover date ~60\% through the training period. For each model set we vary the volume of data by moving the start date of historical training and plot accuracy (Ranking Loss) relative to the model (from within the same model set) with the largest volume of data. The teacher model in all these experiments is same which is a larger model trained on un-sampled data for longer duration than the student models. 

Results as shown in Fig. \ref{fig:chart_cud}, show a significant convergence gap between model sets that widens as the amount of training data decreases. Notably, continuous distillation uses 35\% less data than the later cutover model. %, yet achieves comparable quality.
This suggests that distillation not only accelerates convergence but also that extending distillation over a longer period further enhances this acceleration.

\section{Jointly tuning training data and model size}

A fundamental question to answer while scaling any production model is: \textit{Given a fixed compute budget, how should one trade off model size and the number of training examples \cite{chinchilla}?"} While studied extensively for LLMs and MOEs \cite{chinchilla, kaplan2020scaling, clark2022unified}, with evidence suggesting data scaling is more effective than model scaling for LRMs up to 5B examples \cite{ardalani2022understanding}, this remains under-explored for ad click models trained on over 100B examples \cite{anil2022factory}.

\begin{figure}[ht]
    \centering
    \includegraphics[width=0.9\linewidth]{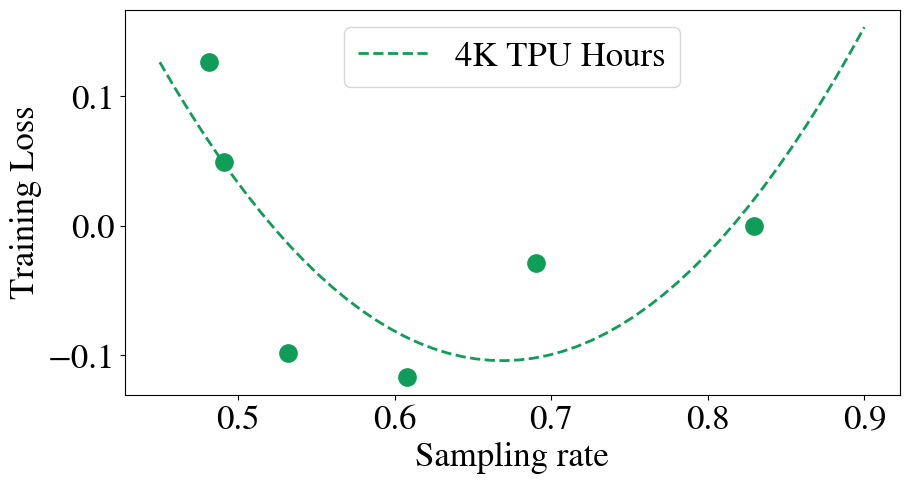}
    \caption{\centering IsoCompute Curve \linebreak {\footnotesize For a fixed training budget of ~4K TPU hours, we varied the sampling rate and model size to achieve optimal quality}} 
    \label{fig:jopt}
\end{figure}

Here we computed the iso-compute curve, where we fixed the compute cost (in terms of TPU \cite{TPU} training time) and varied the sampling rate to control the number of training examples and model size. We observed (in Fig. \ref{fig:jopt}) a similar pattern to Chinchilla's iso-FLOP curves \cite{chinchilla}, where a clear valley in loss exists. This indicates that, for a given compute budget, there is an optimal model size and sampling rate configuration.

\section{Conclusion}
This paper presented strategies and frameworks to enhance the data efficiency of large recommendation models (LRMs). We explored data convergence and approaches to reducing training data, in which continuous downsampling allows for early convergence and  continuous distillation allows for convergence with substantially less data. We also discussed the joint tuning of training data and model size. By adopting these strategies, practitioners can reduce training time, enhance system efficiency, and accelerate model development.

% \bibliography{ref}

\end{document}